\documentclass[aps,twocolumn,prl,amssymb, amsmath,superscriptaddress]{revtex4-2}
\pdfoutput=1 
\usepackage{dcolumn}
\usepackage{bm}
\usepackage[utf8]{inputenc}
\usepackage[T1]{fontenc}
\usepackage[english]{babel}  
\usepackage{euscript}
\usepackage{amssymb,amsmath,amsfonts,amsthm,bm}
\usepackage{mathtools}
\usepackage{verbatim}
\usepackage{enumerate}
\usepackage{tensor}
\usepackage{fancyhdr}
\usepackage{xcolor}
\usepackage{graphicx}
\graphicspath{{pics/}}
\usepackage[colorlinks=true,linkcolor=blue,citecolor=blue,urlcolor=blue]{hyperref}
\usepackage{bbm}
\usepackage{booktabs}

\usepackage{lineno}
\usepackage{subfigure}

\usepackage[percent]{overpic}
\usepackage{dsfont}
\usepackage{physics}

\newcommand{\phii}[0]{\hat{\Phi}}
\newcommand{\mm}[0]{\hat{\mathcal{M}}}
\newcommand{\uu}[0]{\hat{\mathcal{U}}}
\newcommand{\kett}[1]{\left.\ket{#1}\right\rangle}
\newcommand{\braa}[1]{\left\langle\bra{#1}\right.}
\newcommand{\brakett}[2]{\left\langle\braket{#1}{#2}\right\rangle}
\newcommand{\expect}[1]{\mbox{$\langle #1 \rangle$}}
\newcommand{\supoper}[1]{\hat{\mathcal{#1}}}
\renewcommand{\a}[0]{a} 

\begin{document}

\preprint{APS/123-QED}

\title{Observation of metastability in open quantum dynamics of a solid-state system}

\author{Jun-Xiang Zhang}\thanks{These authors contributed equally to this work}
\affiliation{Beijing National Laboratory for Condensed Matter Physics and Institute of Physics, Chinese Academy of Sciences, Beijing 100190, China}
\affiliation{School of Physical Sciences, University of Chinese Academy of Sciences, Beijing 100049, China}
\author{Yuan-De Jin}\thanks{These authors contributed equally to this work}
\affiliation{State Key Laboratory of Superlattices and Microstructures, Institute of Semiconductors, Chinese Academy of Sciences, Beijing, 100083, China}
\affiliation{Center of Materials Science and Opto-Electronic Technology, University of Chinese Academy of Sciences, Beijing 100049, China}
\author{Chu-Dan Qiu}\thanks{}
\affiliation{State Key Laboratory of Superlattices and Microstructures, Institute of Semiconductors, Chinese Academy of Sciences, Beijing, 100083, China}
\affiliation{Center of Materials Science and Opto-Electronic Technology, University of Chinese Academy of Sciences, Beijing 100049, China}
\author{Wen-Long Ma}
\email{wenlongma@semi.ac.cn}
\affiliation{State Key Laboratory of Superlattices and Microstructures, Institute of Semiconductors, Chinese Academy of Sciences, Beijing, 100083, China}
\affiliation{Center of Materials Science and Opto-Electronic Technology, University of Chinese Academy of Sciences, Beijing 100049, China}
\author{Gang-Qin Liu}
\email{gqliu@iphy.ac.cn}
\affiliation{Beijing National Laboratory for Condensed Matter Physics and Institute of Physics, Chinese Academy of Sciences, Beijing 100190, China}
\affiliation{Songshan Lake Materials Laboratory, Dongguan, Guangdong 523808, China}

\date{\today}

\begin{abstract}
Metastability is a ubiquitous phenomenon in non-equilibrium physics \cite{Mori2018,Lin2020} and classical stochastic dynamics \cite{Gaveau1987,Gaveau2000}. It arises when the system dynamics settles in long-lived states before eventually decaying to true equilibria. Remarkably, it has been predicted that quantum metastability can also occur in continuous-time \cite{Macieszczak2016} and discrete-time \cite{Jin2024} open quantum dynamics. However, the direct experimental observation of metastability in open quantum systems has remained elusive. Here, we experimentally observe metastability in the discrete-time evolution of a single nuclear spin in diamond, realized by sequential Ramsey interferometry measurements of a nearby nitrogen-vacancy electron spin. We demonstrate that the metastable polarization of the nuclear spin emerges at around 60,000$\sim$250,000 sequential measurements, enabling high-fidelity single-shot readout of the nuclear spin under a small magnetic field of 108.4 gauss. An ultra-long spin relaxation time of more than 10 s has been observed at room temperature. By further increasing the measurement number, the nuclear spin eventually relaxes into the maximally mixed state.  Our results represent a concrete step towards uncovering non-equilibrium physics in open quantum dynamics, which is practically relevant for the utilization of metastable information in various quantum information processing tasks \cite{Blume-Kohout2008,Blume-Kohout2010}, such as accurate quantum operations \cite{ma2023high, Earnest2018}, quantum channel discrimination \cite{DeBry2023} and quantum error correction \cite{Kang2023}.

\end{abstract}

\maketitle

\emph{Introduction.---}
Metastability, similar to prethermalization in nonequilibrium physics \cite{Mori2018}, often emerges in classical or quantum many-body systems, when the system relaxes into long-lived states before subsequently decaying to true stationarity much slowly. In classical stochastic dynamics, metastability is a manifestation of a separation of time scales that arises from the splitting in the spectrum of the dynamical generator. 

Recently quantum metastability theory has been extended to open quantum dynamics described by Markovian dynamics \cite{Macieszczak2016,Macieszczak2021a,Brown2024}, where the manifold of metastable states is argued to be composed of disjoint states, decoherence-free subspaces, and noiseless subsystems. Such metastability theory is physically interesting from their own right and stimulates the prediction of metastability phenomena in various quantum models \cite{Rose2016,Leboite2017,Gong2018,Gambetta2019,Landa2020a,Flynn2021,Defenu2021,Jager2022,Cabot2022}. It is also of practical importance to utilize the preserved information of quantum processes for quantum information processing \cite{Blume-Kohout2008,Blume-Kohout2010}, especially in the presence of control imperfections or environmental noise. 

 For the continuous-time dynamics described by the Lindblad master equation, metastability can be observed from the distinct two-step decay of temporal correlations \cite{Macieszczak2016}. However, experimental demonstrations remain scare, mainly due to the difficulty of measuring temporal second-order or higher-order correlations with good resolution \cite{Fitzpatrick2017,Letscher2017}. Alternatively, certain signatures of metastability can be observed without measuring the temporal correlations (e.g., through the high-accuracy time-resolved heterodyne detection in the superconducting cavities)  \cite{beaulieu2023}, however, such approaches are generally not applicable to other experimental settings.

  \begin{figure*}
     \centering
     \includegraphics{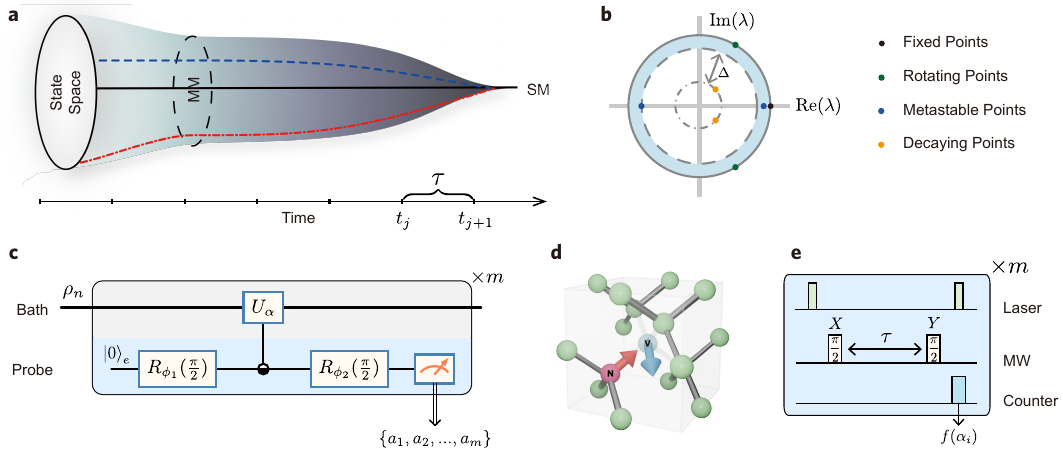}
     \caption{\textbf{Metastability in discrete-time open quantum dynamics and its observation in diamond.}  \textbf{a,} Schematic of metastability in open quantum dynamics. 
     The states outside the metastable manifold (MM, red dot-dashed line) exhibit a two-step relaxation, i.e., first quickly relaxing into the MM and then slowing decaying into the stationary manifold (SM).
      The discrete-time axis implies that the open quantum dynamics is described by sequential quantum channels. \textbf{b,} Spectrum of a quantum channel, with all eigenvalues (solid circles) located within the unit circle of the complex plane. Metastable points emerge in the region $|\lambda|\approx 1$ (labeled with blue ring in the spectrum). The range of metastable region is determined by the spectral gap $\Delta$ between the decaying point with the largest eigenvalue (dot-dashed circle) and the metastable point with the smallest eigenvalue (dashed circle). \textbf{c,} Quantum circuit of sequential RIMs, where $R_{\phi}(\theta)$ is the rotation operator of the probe qubit, $U_{\alpha}$ is a unitary operator of the bath conditioned on the probe state $\ket{\alpha}_e$ ($\alpha=0,1$), and $\{\a_1,...,\a_m\}$ is the sequence of measurement outcomes.  \textbf{d,} The experimental system, with the electron spin of a nitrogen-vacancy (NV) center serving as the probe qubit, and the host $^{14}$N nuclear spin as the quantum bath system. \textbf{e,} Pulse sequences to implement sequential RIMs in experiments. Laser pulses are used to polarize and read out the NV spin state, and resonant microwave pulses are used to manipulate the probe spin state.  
     }
     \label{circuit}
 \end{figure*}

A recent advance in quantum metastability theory is the generalization of the setting from continuous-time to discrete-time open quantum dynamics \cite{Jin2024}. Such a framework describes the more general scenario in which each discrete evolution is induced by a quantum channel (or a completely positive and trace-preserving quantum map) \cite{Watrous2018}, which may not be generated by continuous-time master equations \cite{gudder2008,amato2023}. With this framework, metastable open quantum dynamics of a quantum system can be directly observed by only sequentially measuring an ancilla qubit, instead of continuously measuring the temporal correlations. Remarkably, metastability has been predicted in the commonly-used Ramsey interferometry measurements (RIMs) \cite{Ramsey1950,Lee2002,Degen2017}, which is easy to realize in various quantum platforms. By repeating the RIM on a probe qubit, a nearby bath system can be metastably polarized, which is manifested in statistics of the probe measurement outcomes associated with different quantum trajectories of the bath system. 

Here we report an observation of the metastable open dynamics in diamond based on sequential RIMs, using a nitrogen ($^{14}\mathrm{N}$) nuclear spin as the bath system and the nitrogen-vacancy (NV) electron spin as the probe. Due to the hyperfine coupling between the bath and the probe during free evolution, each round of RIM solely applied on the probe spin induces a quantum channel on the nuclear spin, and sequential RIMs induce discrete-time open quantum dynamics on the nuclear spin (Fig. \ref{circuit}\textbf{a}). By recording the statistical results of the sequential probe measurement, we can continuously monitor the state evolution of the target spin. We directly observe the metastable polarization with varying the number of repetitions $m$, i.e., the bath spin is first steered to metastable (polarized) states for a finite range of $m$, and eventually relaxes towards the true stable (maximally mixed) state as $m$ continues to increase. 


\emph{Metastability in sequential RIMs.---} We illustrate the principle of metastability by considering a typical class of sequential quantum channels on a quantum bath system, with the channel being induced by a probe qubit under a RIM sequence \cite{Jin2024}. 
The probe qubit is coupled to the bath through the general pure-dephasing coupling
\begin{equation}
    \label{Hamilt}
    H = \sigma_e^{z} \otimes B_n+ \mathbb{I}_e \otimes C_n,
\end{equation}
where the subscripts $e$ and $n$ refer to the probe and bath respectively, $\sigma_e^i$ is the Pauli-$i$ operator of the qubit ($i=x,y,z$) with $\sigma_e^z=|0\rangle_{e}\langle0|-|1\rangle_{e}\langle1|$, and $B_n$ $(C_n)$ is the interaction (free) operator of the bath. We denote the qubit rotation along an axis in equatorial plane as $R_{\phi}(\theta)=e^{-i(\cos\phi \sigma_q^x+\sin \phi \sigma_q^y)\theta/2}$, with $\phi$ denoting the rotation axis and $\theta$ the rotation angle.

For a single RIM, the qubit is first initialized to state $|0\rangle_e$, and then prepared in a superposition state $|\psi\rangle_e=(|0\rangle_e-i e^{i\phi_1}|1\rangle_e)/\sqrt{2}$ by a rotation $R_{\phi_1}(\frac{\pi}{2})$. After the composite system evolving with the pure-dephasing coupling [Eq. (\ref{Hamilt})] for time $\tau$, the qubit undergoes another rotation $R_{\phi_2}(\frac{\pi}{2})$, and is finally projectively measured in the basis of $\sigma_e^z$ with the measurement outcome $\a\in\{0,1\}$. 

The RIM sequence of the probe spin induces a quantum channel $\Phi$ on the bath, 
\begin{equation}
    \label{channel}
    \Phi(\rho_n)={\rm Tr}_{e}\left[U(\rho_e\otimes \rho_n) U^{\dagger}\right]=\sum_{\a=0,1}M_{\a}\rho_n M_{\a}^{\dagger},
\end{equation}
where $\rho_e=|\psi\rangle_e\langle \psi|$ ($\rho_n$) is the initial state of the probe (bath) spin, $U=e^{-iH\tau}$ is the propagator for the composite system. By partially tracing over the probe spin, we obtain the Kraus operator $M_{\a}=[U_0-(-1)^{\a}e^{i\Delta\phi}U_1]/2$ with $U_{\alpha}=e^{-i[(-1)^{\alpha} B_n+ C_n]\tau}$ and $\Delta\phi=\phi_1-\phi_2$ being the phase difference between the rotation axes of two $\pi/2$ pulse in RIMs. While obtaining the outcome $\a$, the bath state is steered to $M_{\a} \rho_n M_\a^\dagger$ with probability $p(\a)=\Tr(M_{\a} \rho_n M_\a^\dagger)$. Note that $\sum_\a M_\a^\dagger M_{\a}=\mathbb{I}$ ensures that $\sum_{\a}p(\a)=1$. 

The behaviors of repetitive quantum channels can be clearly revealed by the spectral decomposition of a single channel (see Method). A quantum channel has at least one fixed point (Fig.\ref{circuit}\textbf{b}), which is the state that remains unchanged after the channel \cite{Arias2002,Albert2019}.  It has been proved that the fixed points depend on the commutativity of $B_n$ and $C_n$ (see Method). If $B_n$ commutes with $C_n$, i.e., $[B_n,C_n]=0$, the fixed points are the simultaneous eigenstates of $B_n$ and $C_n$, then sequential RIMs of the probe spin will polarize the bath system to one of the eigenstates, also constituting a quantum non-demolition (QND) measurement on the bath spin \cite{Bauer2011,Ma2023,Wang2023}. However, if $[B_n,C_n]\neq0$, the fixed points are the maximally mixed state in the whole space (or subspace) of the bath, then sequential RIMs cannot effectively measure the bath system but only perturb it. 

Quantum metastability emerges when $[B_n,C_n]\neq0$, but $C_n$ is a small perturbation of $B_n$, then the bath will be polarized for a finite range of repetitions of RIMs. We suppose that the channel has $r$ fixed points and $q-r$ metastable points, which collectively span a $(q-1)$-dimensional metastable manifold (MM). The metastable region of the polarization of the whole state space can be estimated as $(1-|\lambda_{q+1}|)^{-1}\ll m\ll (1-|\lambda_q|)^{-1}$ when $\lambda_q$ is very close to 1, where $\lambda_q$ is the eigenvalue of metastable point with the smallest eigenvalue. As $m$ increases and enters the metastable region, the system’s state evolves into the MM. Beyond this region, the system further relaxes into the stationary manifold (SM), which is spanned by the fixed points. This process, where the system first reaches the MM and then relaxes to the SM, is referred to as a two-step relaxation (see Fig.\ref{circuit}\textbf{a}).

The metastability behaviors in sequential RIMs can be directly monitored by the measurement statistics of the probe spin \cite{Jin2024}. To see this, we decompose the average dynamics of sequential quantum channels into stochastic trajectories,
\begin{equation}
    \Phi^m(\rho_n)=\sum_{\a_1,...,\a_m=0}^1 \mathcal{M}_{\a_m}\cdots \mathcal{M}_{\a_1}(\rho_n),
\end{equation}
where $\mathcal{M}_{\a_i}(\cdot)=M_{\a_i}(\cdot)M_{\a_i}^\dagger$ with $\a_i\in\{0,1\}$ being the measurement outcome of the $i$th RIM. For a trajectory with the sequence of measurement outcomes $\{\a_1,...,\a_m\}$, the bath is steered to $\rho'_n=\mathcal{M}_{\a_m}\cdots \mathcal{M}_{\a_1}(\rho_n)/p(\a_1,...,\a_m)$ with probability $p(\a_1,...,\a_m)=\Tr[\mathcal{M}_{\a_m}\cdots \mathcal{M}_{\a_1}(\rho_n)]$.
For a trajectory with $\{\a_1,...,\a_m\}$ with the number of outcome $0/1$ being $m_0/m_1$ ($m_0+m_1=m$), we define the measurement polarization $X=(m_0-m_1)/(2m)$, denoting the different classes of stochastic trajectories that the target system undergoes. The measurement distribution of $X$ can show several peaks, with each peak corresponding to the quantum trajectories that lead to a fixed point (or metastable state) of the channel (see Methods). So in the metastable region, the distribution exhibits up to $q$ peaks, corresponding to the metastable states. However, as $m$ surpasses this region, the number of peaks gradually reduces to $r$, reflecting the final stationary states of the bath.

\emph{Experimental implementation in the NV system.---}
We demonstrate this protocol experimentally on an NV center and a nearby nuclear spin in a high-purity diamond  (Fig.\ref{circuit}\textbf{d}).
Nuclear spins around NV centers have great potential for building quantum networks \cite{network2013}, storing quantum information \cite{register2012}, establishing quantum simulator \cite{simulator2013,fault2022,crystal2021}, sensing inertial parameters \cite{sensing2021}, and so on. Access to the nuclear spins is usually based on their hyperfine interaction with the electron spin and the optical interface of the NV center. It is therefore an interesting and fundamental problem to monitor and understand the dynamics of these nuclear spins under multiple measurements acting on the central electron spin.
  \begin{figure*}[t]  
    \centering
    \includegraphics{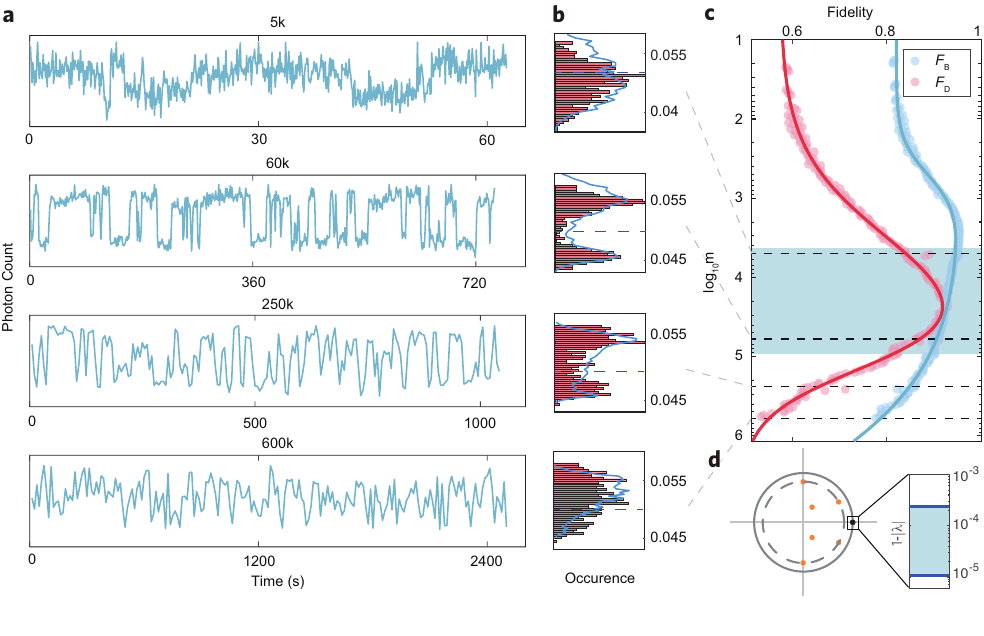}  
    \caption{\textbf{Metastable dynamics of a $^{14}$N nuclear spin induced by sequential RIMs of a nearby NV electron spin.} 
 \textbf{a-b,} Typical PL time trace of an NV center under sequential RIMs and distribution of consecutive measurements. The duration of free evolution, $\tau= 374 $ ns. The number of measurement repetition, $m$, is 5 k, 60 k, 250 k, 600 k, from top to bottom, respectively. At a relatively small $m = 5\,$k, it is difficult to distinguish the nuclear spin quantum states. Increasing $m$ to $60\,$k results in the jump signal being concentrated in two distinct photon count intervals.
  As $m$ increase further to $250\,$k, the overlap between the two distribution peaks becomes more pronounced. When $m$ exceeds the metastable region, the photon counting peaks gradually merge as $m$ continues to increases. The blue envelopes indicate the normalized numerical simulation results. \textbf{c,} The evolution of fidelity for quantum trajectories in the numerical simulation of sequential RIMs. The initial thermal state evolves through metastable dynamics: trajectories are polarized to either the dark state ($|1\rangle_n$) or the bright state, represented by the subspace spanned by $|0\rangle_n$ and $|-1\rangle_n$. The fidelity $F_{\rm D,B}$ quantifies how close the quantum states represented by the trajectories are to the ideal dark or bright states, respectively. 
  \textbf{d,} The channel spectrum, with one fixed point, two metastable points and six decaying points, the absolute value of metastable points are shown in right.  The metastable region after the relaxation of $\lambda_3$ and before that of $\lambda_2$ is labeled with blue shade in \textbf{c} and \textbf{d}. Monte Carlo simulations are conducted with a tilt angle $\theta=8.8^\circ$ and 3000 samples. 
   }
    \label{data}
\end{figure*}

Specifically, an NV electron spin is a spin-1 system with a zero-field splitting $D=2.87$ GHz between its $|0\rangle_e$ ($m_s=0$) and $|\pm1\rangle_e$ ($m_s=\pm1$) state. In our experiment, a moderate external magnetic field of $B=108.4 \pm 0.2$ G is applied along the NV symmetry axis ($z$ axis) to lift the degeneracy of the $|\pm1\rangle_e$ states and we work in the $\{|0\rangle_e, |-1\rangle_e\}$ subspace. 
The Hamiltonian in Eq. \eqref{Hamilt} becomes (in the rotating frame)
\begin{equation}
    \label{HamiltNV}
    H = A_{zz}S_{z}I_z+Q I_z^2-\gamma_n\bm{B} \cdot \bm{I}+\tilde H_1,
\end{equation}
where $S_{z}\ (I_{z})$ is Pauli-$z$ operator for the NV electron spin ($^{14}\mathrm{N}$ nuclear spin) with $\bm{I}=(I_x, I_y, I_z)$, $A_{zz}=-2.16$ MHz ($A_\perp=-2.63$ MHz) is the $zz$-component (transverse component) of the hyperfine tensor, $Q=-4.95$ MHz is the quadrupolar splitting, $\gamma_{n}$ ($\gamma_{e}$) being the gyromagnetic ratio of $^{14}\mathrm{N}$ nuclear spin (electron spin),
and $\tilde H_1=\sum_\alpha|\alpha\rangle_e\langle \alpha|\otimes H^\alpha_n$ is a second-order perturbation term with $    H_n^{\alpha}\approx\frac{\gamma_e(2-3|\alpha|)}{2D}[-\gamma_e(B_x^2+B_y^2)+2A_\perp(B_xI_x+B_yI_y)]$.
To introduce a non-commutativity between $B_n=A_{zz}I_z$ and $C_n=Q I_z^2-\gamma_n\bm{B} \cdot \bm{I}+\tilde H_1$,  the external magnetic field is slightly misaligned with the NV axis. Under this circumstance, the perturbation terms $-\gamma_n (B_xI_x+B_yI_y)$ and $\tilde{H}_1$ lead to the metastable polarization of $^{14}\mathrm{N}$.


For each RIM, the resulting quantum state of the NV spin is read out by counting its spin-dependent photon number. As shown in Fig. \ref{circuit}\textbf{e}, a 532-nm laser is used to excite the NV center and all experiments are performed at room temperature. Due to the low photon emission rate of the single NV center and the short readout time (200 ns, see discussion below), on average only 0.05 photons can be detected in each measurement. To obtain a reasonable signal-to-noise ratio, we simply sum the counts of many measurements. The relatively low readout efficiency of the NV electron spin corresponds to a weak measurement of the electron spin (see Methods), which make it more challenging to observe metastable polarization of the nuclear spin.
We anticipate that such metastable polarization can be more efficiently realized with projective (strong) measurement of the NV electron spin \cite{2014NP_backaction,bonato2016}. 

Figure~\ref{data} presents the metastable dynamics of the target nuclear spin under sequential RIMs acting on the NV electron spin. Before the measurement, the $^{14}\mathrm{N}$ nuclear spin is in the thermal state, which means that its population is evenly distributed among the set of states $\{|0\rangle_n,|\pm1\rangle\}_n\}$ ($m_I={0,\pm1}$). 
For a small number of RIMs ($m < 5\,$k), the shot-noise of the photon counts exceeds the count difference of the states to-be resolved (the signal), so that no signal can be detected in this range. 
As the number of RIMs increased, the measurement results begin to concentrate on two regions,   
a clear signature of the metastable polarization of the bath spin, as can be seen in the middle slice of Fig.~\ref{data}\textbf{a},\textbf{b}. 
 We classify the trajectories based on the average photon number with a threshold of 0.05. If the average photon number is less than 0.05, it is classified as a $dark$ state, otherwise as a $bright$ state.
Since the $^{14}\mathrm{N} $ nuclear spin is a three-level system, three peaks are expected, and the feature of only two peaks indicates that two of them cannot be distinguished through current weak measurements of the electron spin.
Through numerical simulations (see Extended Data Fig.~\ref{fig:edf2} \textbf{c}), we find that three peaks are observed at about 5 k measurements and two of them are depolarized before 60 k measurements.
For the measurement number of  ($m > 5\,$k), the experimentally resolved bright state corresponds to the projector of the subspace spanned by $ |0\rangle_n $ and $ |-1\rangle_n $. This projector, together with $|1\rangle_n$ (dark state), are the extremely metastable states (EMSs) of the one-dimensional MM spanned with the fixed point and the metastable point with higher eigenvalue of the quantum channel. 
Beyond the metastable region ($m > 600\,$k), the two peaks mentioned above gradually disappear and a single peak appears, which corresponds to the stable state of the system. More data can be found in Extended data Figs.~\ref{fig:Ex_data1} and \ref{fig:edf2}.

 To quantitatively understand the metastable dynamics of the nuclear spin, we perform numerical simulations that take into account the parameters of hyperfine coupling, the strength of the external magnetic field, and the readout efficiency of the NV electron spin. 
 The metastable states can be quantified by comparing them with the ideal polarization states, $\rho_{\rm D}=|1\rangle_n\langle 1|$ for the $dark$ state and the projector $\rho_{\rm B}=(|0\rangle_n\langle 0| + |-1\rangle_n\langle -1|)/2$ for the $bright$ state. In  particular, their evolution fidelities are calculated as  $F_{\rm D(B)} = F(\rho_{\rm D(B)}, \rho_n)$, where the fidelity $F(\rho, \sigma) = \Tr\sqrt{\rho^{1/2} \sigma \rho^{1/2}}$ and $\rho_n$ is the state of the specific trajectory. As shown in Fig.~\ref{data}\textbf{c}, the experimentally observed metastable dynamics of the $ ^{14}\mathrm{N}$ can be well reproduced with a tilted angle of the magnetic field  of $\theta=8.8^\circ$.
 By choosing a smaller angle of the external magnetic field, the metastable stage of all three polarized nuclear spin states with $m_I={0, \pm1}$ can last for a long time, which enabling the observation of three-peak jumps in the experiment, see Extended Data Fig.~\ref{edf3} for details. 

\begin{figure}
     \centering
     \includegraphics[width=8.5cm]{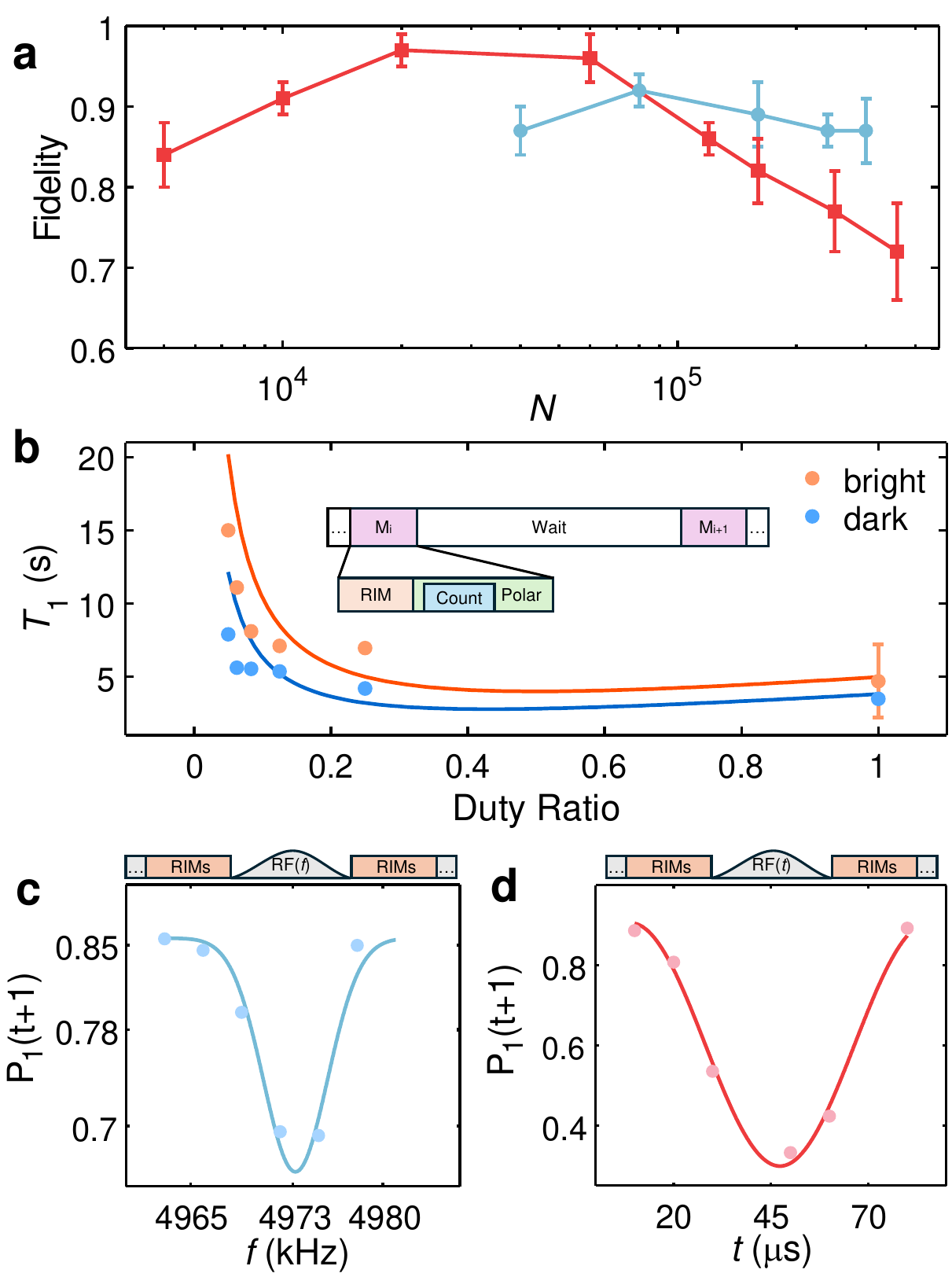}
     \caption{\textbf{Single-shot readout of the nuclear spin state.} 
     \textbf{a,} Readout fidelity of the $^{14}$N nuclear spin as a function of the measurement repetition number. The data points with $\theta=7^\circ\pm2^\circ$, $\tau= 374 $ ns are shown in red squares, while the data points with $\theta=7^\circ\pm2^\circ$, $\tau= 370 $ ns are shown in blue circles. 
     \textbf{b,} $^{14}$N nuclear spin relaxation time (bright state in orange and dark state in blue) as a function of the duty cycle of the excitation laser pulses. The inset illustrates the experimental pulse sequence. The duty cycle is defined as the ratio of the measurement time to the waiting time in the sequence.
     \textbf{c,} $^{14}$N nuclear spin NMR, obtained by  varying the radio-frequency (RF) pulse frequency at a fixed pulse duration of 30 $\mu$s. 
     \textbf{d,} $^{14}$N nuclear spin Rabi oscillations. A resonant RF pulse of a specific length is applied between every two sequential RIMs to flip the nuclear spin between the $|0\rangle_n$ and $|+1\rangle_n$ states, while the electron is at the $|0\rangle_n$ state.
     }
     \label{data1}
 \end{figure}

\emph{Single-shot readout of a single nuclear spin---}
The observed metastability provides a simple and efficient way to realize high-fidelity single-shot readout of the nuclear spin. 
Under a magnetic field tilted angle $\theta=7^\circ\pm2^\circ$ and a RIM interval of $\tau= 374 $ ns, the quantum jumps of the nuclear spin are clearly observed in the measurement number range from  5 k to 250 k. We then characterize the readout fidelity of the nuclear spin using the method developed in \cite{2013drean, Lucas2010} , and the results are summarized in Fig.~\ref{data1}\textbf{a}.
It is worth noting that both the external magnetic field and the RIM interval affect the occurrence of the metastable states, so as to the optimal readout region, as shown in Fig.~\ref{data1}\textbf{a}.
With a repetition number of $m = $ 80 k (total measurement time 1.73 s), a readout fidelity of $97\%\pm2\%$ with a threshold count value of 2950 is achieved.


Next, we investigate the relaxation mechanism of the $^{14}$N nuclear spin under repeated measurements. Previous studies have shown that the flip-flop process in the  excited state of the NV center is the dominant factor, and usually a large external magnetic field (> 2000 G) is applied to suppress this term so that single-shot readout of the nuclear spin can be achieved  \cite{2013drean,2010single}. Alternatively, our results show that the nuclear spin relaxation can be largely suppressed by simply shortening the duration of the excitation laser. Moreover, by inserting additional waiting intervals between the sequential RIMs, a longer nuclear spin $T_1$ is observed, as summarized in  Fig.~\ref{data1}\textbf{b}. The longest $T_1$ measured in our experiment is 15 s. 

Finally, we perform nuclear magnetic resonance (NMR) and Rabi measurements to demonstrate the coherent control of the $^{14}$N nuclear spin. The nuclear spin state is first prepared by a RIM measurement sequence (either in the bright or dark state), then a  RF pulse is applied to flip the nuclear spin, and a second RIM measurement is performed to determine the nuclear spin state. The RF pulse is either fixed in duration (30 $\mathrm{\mu}$s for NMR) or fixed in frequency (resonant frequency for Rabi). The sequence are repeated for  80 k times and the probability of nuclear spin flip are recorded. As mentioned above, the experimentally resolved bright state corresponds to the maximally mixed state in the subspace $\{|0\rangle_n $, $ |-1\rangle_n\}$, so we fit the fluorescence time trace using a three-state hidden Markov model.  Fig.~\ref{data1}\textbf{c} shows the NMR signal at 4973.0 kHz, which corresponds to the $^{14}$N-bath nuclear spin being in $|+1\rangle_n$ state. Fig.~\ref{data1}\textbf{d} shows the Rabi oscillation of the $^{14}$N nuclear spin between the states $|+1\rangle_n$ and $|0\rangle_n$. These results further confirm the observed quantum jump signal originating from the $^{14}$N nuclear spin.

\emph{Conclusion and outlook.---}
In conclusion, we have directly observed metastability in the discrete-time open quantum dynamics of a single nuclear spin in diamond, induced by sequential RIMs of a probe NV electron spin. The observed metastable polarization of the nuclear spin is of practical importance for constructing QND measurements of a quantum system that can be easily extended to other physical platforms, including trapped-ions, superconducting circuits, and other semiconductor single-spin systems. On the other hand, our results also point to a potential issue in constructing quantum networks with NV centers or other single-spin systems, since sequential measurements of an ancilla can cause its nearby memory to  equilibrate towards a maximally mixed state, leading to the loss of information about the stored state.

The generality of our protocols is also a blueprint for future studies of non-equilibrium phenomena in open quantum systems. It will be interesting to realize more general channels in both theory and experiments in the future, whose fixed points may include decoherence-free subspaces or noise-free subsystems, and to explore the metastability phenomena for protected coding and logical gates. Such phenomena can be directly relevant to quantum information processing, since the fixed points of a quantum channel with preserved information can become metastable when control errors or environmental noise are taken into account. 



\appendix
\bibliography{refer}

\textbf{Acknowledgments}. This work is supported by the Innovation Program for Quantum Science and Technology (Grants Nos. 2023ZD0300600, 2021ZD0302300), the National Key Research and Development Program of China (Grants No. 2023YFA1608900), the National Natural Science Foundation of China (Grant Nos. 12022509, 11934018, T2121001), and the Chinese Academy of Sciences (Grant Nos. YSBR90, YSBR-100, E0SEBB11, E27RBB11).


\textbf{Author contributions}. W.L.M and G.Q.L. conceived the project, J.X.Z conducted the experiments with the supervision of G.Q.L, Y.D.J and C.D.Q performed theoretical analysis and simulations with the supervision of W.L.M. All authors contributed to the writing of the paper.

\newpage\
\textbf{Methods}

\textbf{Experimental setup and sample information}.

The experiments are performed with  a [100]-oriented, type IIa single-crystal diamond produced by Element Six, with a natural $^{13}$C isotopic abundance of 1.1\%. To improve the phonton collection efficiency, a solid immersion lens (SIL) is etched onto the diamond surface. Under saturation conditions, the photon counting rate of the NV center is about 300 kcps. The diamond was mounted on a custom-built confocal microscope, and a 532-nm laser is used to initialize and read out the NV spin states. The laser pulses are generated with an acoustic optical modulator (AOM, Gooch $\&$ Housego 3350-199). The fluorescence emitted from the NV center is  collected with the same objective, passed through a dichroic mirror and a 650-nm long-pass filter, then coupled into a multimode fiber and detected with a single photon detector (SPD). The fluorescence photons are counted with an NI-data acquisition card (NI-6343).

To control the NV electron spin and the $^{14}$N nuclear spins, we use an arbitrary waveform generator (AWG, Techtronix 5024C) to generate transistor-transistor logic (TTL) signals and low-frequency analog signals. The synchronized TTL signals are used to control the AOM, RF switches and counters. For electron spin manipulation, the microwave (MW) carrier signal from a MW source (Rohde $\&$ Schwarz SMIQ03B) is combined with two analog AWG signals using an IQ-mixer (Marki Microwave IQ1545LMP). An additional analog signal is used to generate the RF signal for nuclear spin manipulation. Both the MW and RF signals are amplified by amplifiers (Mini Circuits, ZHL-16W-43-S+ and 32A-S+). The MW and RF signals are transmitted to the NV position via a gold co-planar waveguide deposited on the diamond surface.

As can be seen in the ODMR spectrum of the NV center, there is no splitting due to hyperfine interaction with strongly coupled ${}^{13}\mathrm{C}$ nuclear spin, and a static magnetic field $B$ = 108.4 $\pm$ 0.2 G is applied close to the NV symmetry axis. The spin transition between the $|0\rangle_e\leftrightarrow|-1\rangle_e$ electronic spin states is addressed via microwave pulses. Figure S2 shows the free-induction decay (FID) signal of the measured NV center, which gives $T_{2e}^{*} = 2.9$ $\mathrm{\mu}$s. 

For the $^{14}$N nuclear spin, the measurement strength oscillates as a function of the RIM duration $\tau$. Considering the nuclear spin relaxation and the NV electron spin dephasing time, $\tau=$ 374 ns is chosen to perform the experiments. Under these circumstances, the relaxation induced by sequential RIMs precedes the intrinsic $^{14}$N nuclear spin $T_1$ process. For each RIM, an optical pulse of 220 ns is used to read out the population of the NV electron spin. After the first measurement, the nuclear spin randomly collapses to |0⟩${}_n$ or |$\pm$1⟩${}_n$, and in the following measurement, it remains at the same state until a quantum jump occurs.

\textbf{Quantum channels and their representations}

Quantum channel, also called completely positive and trace-preserving (CPTP) map, describes the most general (closed or open) quantum dynamics that a quantum system can undergo. Each quantum channel has four different representations: the Kraus representation, the Stinespring representation, the natural representation, and the Choi representation. The Kraus representation is the most commonly used and characterizes a channel by a set of Kraus operators $\{M_\a\}_{\a=1}^r$, which satisfy $\sum_\a M^{\dagger}_{\a}M_{\a} =\mathbb{I}$, such that $\Phi(\cdot)=\sum_{\a=1}^r M_{\a} (\cdot) M_{\a}^\dagger=\sum_{\a=1}^r \mathcal{M}_{\a}(\cdot)$. The Stinespring representation is a dilation of a quantum channel, which is realized by coupling the bath system to a probe system, subjecting the composite system  to a unitary evolution and then tracing over the probe system, i.e. $\Phi(\rho_n)=\Tr_e[U(\rho_e\otimes\rho_n)U^\dagger]$, where ${\rm Tr}_e[\cdot]$ denotes the partial trace over the probe qubit.

Since a quantum channel acts on the operator space of the bath system and can be considered as a superoperator, it is more convenient to use its natural representation in the Hilbert-Schmidt (HS) space. In the HS space, an operator is transformed into a vector ($X=\sum_{i,j=1}^d x_{ij}|i\rangle\langle j|\leftrightarrow |X\rangle\rangle=\sum_{i,j=1}^d x_{ij}|ij\rangle\rangle$), so that each superoperator is transformed into a single matrix ($X(\cdot)Y\to X\otimes Y^{T}$). Thus, the natural representation of $\Phi$ is $\hat{\Phi}=\sum_{\a=1}^r \hat{\mathcal{M}}_\a$ with $\hat{\mathcal{M}}_\a=M_\a\otimes M^*_\a$. Note that we add hats on the operators of the HS space.

In the natural representation, we can spectrally decompose the quantum channel as
\begin{equation}
  \hat\Phi=\sum_i \lambda_i |R_i\rangle\rangle \langle\langle L_i|,
\end{equation}
where $\lambda_i$ is the $i$th eigenvalue with the corresponding right (left) eigenvector$\kett{R_i}\,(\kett{L_i})$, satisfying $\hat \Phi\kett{R_i}=\lambda_i \kett{R_i}$, $\hat \Phi^\dagger\kett{L_i}=\lambda_i^{*} \kett{L_i}$, and the biorthonormalization condition $\brakett{L_i}{R_j}=\mathrm{Tr}(L_i^\dagger R_j)=\delta_{ij}$ with $\delta_{ij}$ being the Kronecker delta. 
The eigenvalues $\{\lambda_i\}$ of a quantum channel are all located within a unit disk of the complex plane. The eigenspaces with $\lambda=1$ are called \textit{fixed points}, those with $\lambda=e^{i\varphi}$ and $\varphi\neq 0$ are rotating points, and those with $|\lambda|<1$ are decaying points such that the information of such points collapses as $\phii^m\kett{R_j}=\lambda_j^m\kett{R_j}\to 0$ for $|\lambda_j|<1$. 

\textbf{Metastability in sequential quantum channels}

Of particular interest are the decaying points with eigenvalue $|\lambda_i|\approx 1$, which are called \textit{metastable points}, and are denoted by $\kett{\rho_{\rm fix}^i}$. Quantum metastability emerges when there are metastable points in the channel spectrum. Although the information of these metastable points will be finally lost, it can be well preserved within a certain range of $m$. Specifically, for the channel with $r$ fixed points and $q-r$ metastable fixed points, the contributions of the metastable points cannot be neglected when $m\ll \mu' = 1/\big|\ln|\lambda_q|\big|$, while the contribution of the other decaying points decays fast as $m$ grows and can be omitted when $m\gg \mu''=1/\big|\ln|\lambda_{q+1}|\big|$. So $\mu'$ and $\mu''$ delimit a metastable region:
\begin{equation}
  \frac{1}{\big|\ln|\lambda_{q+1}|\big|}\ll m\ll \frac{1}{\big|\ln|\lambda_q|\big|}.
\end{equation}
When the eigenvalue of metastable points are very close to 1, and the decaying points relax fast, which is suitable for our case, the metastable region can be estimated as 
\begin{equation}
    m\ll\frac{1}{|\ln|\lambda_q||}\approx\frac{1}{1-|\lambda_q|}
\end{equation}

In the metastable region, $\lambda_i^m\approx 1$ for $i\leq q$ and $\lambda_j^m\approx 0$ for $j>q$, then  we have
\begin{equation}
  \phii^m\kett{\rho}\simeq\sum_{i=1}^r c_i\kett{\rho_{\rm fix}^i}+\sum_{j=r+1}^{q}\tilde c_j \kett{R_j},
\end{equation}
for initial state $\kett{\rho}=\sum_{i} c_i \kett{R_i}$ with $\tilde c_j=c_j e^{im\varphi_j}$. We note that the metastable points, as decaying points, are trace-zero, so any metastable state in metastable manifold is a superposition of both fixed points (trace-1) and metastable points, denoted by $\{c_1,...,c_{r},\tilde c_{r+1},...,\tilde c_q\}$, with $\sum_{i=1}^r c_r=1$, corresponding to a point in the $(q-1)$ dimensional HS subspace.

We can transform from the HS subspace to the metastable manifold (MM), which contains the metastable states as a convex combination of $q$ extreme metastable states (EMSs)
\begin{equation}
  |\rho_{\rm MS}\rangle\rangle=\sum_{\nu=1}^q p_\nu |\rho_{\rm EMS}^\nu\rangle\rangle
\end{equation}
where $p_\nu=\brakett{P_\nu}{\rho_{\rm MS}}$, $P_\nu$ is the projector to $\kett{\rho_{\rm EMS}^\nu}$, i.e. $\langle\langle{ P_\nu}|{\rho_{\rm EMS}^\mu}\rangle\rangle=\delta_{\nu\mu}$ and $\sum_\nu { P_\nu}=\mathbb{I}$, thus $\sum_{\nu}p_\nu=1$, and EMSs are extremal points of the convex hall.

For the case of $r=1$ and $q=2$, i.e. the channel has one fixed point and one metastable point, forming a one-dimensional MM with the EMSs
\begin{equation}
    |\rho^{1,2}_{\rm EMS}\rangle\rangle=|\rho_{\rm fix}\rangle\rangle+c_2^{M,m}|{R}_2\rangle\rangle/h,
    \label{1DEMS}
\end{equation}
where $c_2^{M}$ ($c_2^{m}$) is the maximal (minimal) eigenvalue of $L_2$, and $h=\sqrt{\brakett{L_2}{L_2}\brakett{R_2}{R_2}}$ is a normalization coefficient.

\textbf{Metastability in sequential RIMs}

The dynamics of the bath system after a RIM sequence of the probe is described by a quantum channel, which can be represented in the Stinespring representation as
\begin{equation}
    \Phi(\rho_n)=\Tr_e[U(\rho_e\otimes\rho_n)U^\dagger]
\end{equation}
where $\rho_e=R_{\phi_1}(\frac{\pi}{2})\ket{0}_e\bra{0}R^\dagger_{\phi_1}(\frac{\pi}{2})$.
For the Hamiltonian 
\begin{equation}\label{H}
    H=\sigma^z_e\otimes B_n+\mathbb{I}_e\otimes C_n,
\end{equation}
the unitary evolution can be decomposed as $U=e^{-iH\tau}=\sum_{\alpha=0,1}|\alpha\rangle_e\langle\alpha|\otimes U_{\alpha}$ with $U_{\alpha}=e^{-i[(-1)^{\alpha} B_n+ c_n]\tau}$. After taking the trace over the probe, the channel can be transformed to the Kraus representation as 
\begin{equation}
    \Phi(\rho_n)=\sum_{\a=0,1} \mathcal{M}_{\a}(\rho_n)=\sum_{\a=0,1}M_{\a}\rho_n M_{\a}^{\dagger},
\end{equation}
where $\mathcal{M}_\a(\cdot)=M_{\a} (\cdot) M_{\a}^{\dagger}$ is a superoperator with the Kraus operator $M_{\a}=[U_0-(-1)^{\a}e^{i\Delta\phi}U_1]/2$ and $\Delta\phi=\phi_2-\phi_1$ is the phase difference between the rotation axes of $R_{\phi_1}(\frac{\pi}{2})$ and $R_{\phi_2}(\frac{\pi}{2})$. 
Then the natural representation of the channel for the RIM case is
\begin{equation}
    \phii=\mm_0+\mm_1=(\uu_0+\uu_1)/2,
\end{equation}
where $\mathcal{\hat U_\alpha}=U_\alpha\otimes U_\alpha^*$, and $\mathcal{\hat M}_\a=M_\a\otimes M_\a^*$.

We have proved that the spectrum of the quantum channel induced by RIM is determined by the commutativity of $B_n$ and $C_n$. If $[B_n,C_n]=0$, then $B_n$ and $C_n$ can be diagonalized simultaneously
\begin{equation}
    B_n=\sum_{k=1}^d b_k |k\rangle\langle k|,\quad C_n=\sum_{k=1}^d h_k |k\rangle\langle k|.
\end{equation}
The fixed points are spanned by all the the rank-1 projectors $\{P_k\}_{k=1}^d$. While if $[B_n,C_n]\neq0$, $B_n$ and $C_n$ can only be block diagonalized with a unitary transformation,
\begin{equation}
  B_n=W\left(\bigoplus_{j=1}^r B_{n,j}\right) W^{\dagger}, \quad C_n=W\left(\bigoplus_{j=1}^r C_{n,j}\right) W^{\dagger}
\end{equation}
where $r< d$ is the number of blocks. Such a block diagonalization partitions the Hilbert space of the bath system into a direct sum of $r$ subspaces $\mathcal{H}=\bigoplus_{j=1}^r\mathcal{H}_j$. Then we prove that the fixed points in this case are spanned by all the projectors $\{\Pi_j\}_{j=1}^r$ to the blocks. Note that since $\rank(\Pi_j)\geq 1$ and $r<d$, the number of bases for fixed points in the latter case is smaller than that of the former case. The decreasing of fixed points indicates the growth of decaying points. 

However, we have proved that when $C_n$ is a perturbation of $B_n$, i.e. $||C_n||/||B_n||$ is small, the emerged decaying points are actually metastable points and the EMSs of this case are approximately the eigenstates of $B_n$ \cite{Jin2024}.

\textbf{Model for the NV system}

The full Hamiltonian for the system composed of the NV electron and the $^{14}$N nuclear spin is 
\begin{equation}
    H=DS_z^2+\gamma_e \vb*{B}\cdot\vb*{S}+\vb*{S}\cdot\mathbb{A}\cdot\vb*{I}+QI_z^2+\gamma_n \vb*{B}\cdot\vb*{I},\label{HamiltFull}
\end{equation}
where $\mathbb{A}$ is the hyperfine interaction tensor, containing only the $zz$-component $A_{zz}S_zI_z$ and the transverse component $A_{\perp} (S_xI_x+S_yI_y)$ for the $^{14}$N nuclear spin.
Since the large zero-field splitting of the NV electron spin suppresses the transition between $|m_s=0\rangle$ and $|m_s=\pm 1\rangle$, the Hamiltonian Eq. \eqref{HamiltFull} can be well approximated as a pure-dephasing form in Eq. \eqref{Hamilt} under the second-order perturbation \cite{Childress2006}. 

Let $H_0=DS_z^2-\gamma_eB_zS_z+S_zA_{zz}I_z+QI_z^2-\gamma_n \vb*{B}\cdot \vb*{I}$ and $H_1=-\gamma_e(B_xS_x+B_yS_y)+A_\perp(S_xI_x+S_yI_y)$, then $H_1$ is a perturbation on $H_0$. The second order perturbation gives $\tilde H_1=\sum_\alpha|\alpha\rangle_e\langle \alpha|\otimes H^\alpha_n$, in which
\begin{equation}
\begin{aligned}
            H_n^\alpha&=P_\alpha^e H_1 \frac{1}{E_\alpha-(\mathbb{I}_e-P_\alpha^e)H_0(\mathbb{I}_e-P_\alpha^e)}H_1P_\alpha^e,
\end{aligned}
\end{equation}
with $E_\alpha=D\alpha^2-\gamma_e B_z\alpha$.
Specifically, we have
\begin{equation}
    H_n^{\alpha}\approx\frac{\gamma_e(2-3|\alpha|)}{2D}[-\gamma_e(B_x^2+B_y^2)+2A_\perp(B_xI_x+B_yI_y)].
\end{equation}
In the rotating frame associated with $H_R=DS_z^2+\gamma_e B_zS_z$, the Hamiltonian becomes 
\begin{equation}\label{Hnv}
\begin{aligned}
        H&=e^{iH_Rt}(H-H_R)e^{-iH_Rt}\\
        &=A_{zz}S_{z}I_z+Q I_z^2-\gamma_n\bm{B} \cdot \bm{I}+\tilde{H_1},
\end{aligned}
\end{equation}
which is time-independent since $[H_R,\tilde H_1]=0$ under the second-order perturbation. The above equation can be cast in the form of Eq. \eqref{H} with $B_n=A_{zz}I_z$ and $C_n=Q I_z^2-\gamma_n\bm{B} \cdot \bm{I}+\tilde{H_1}$.

When the magnetic field is perfectly aligned along the NV axis, $[B_n,C_n]=0$, the fixed points of the channel are $P^n_\beta=|\beta\rangle_n\langle \beta|$, with $\beta=0,\pm 1$. While when the magnetic field deviates slightly from the axis, $[B_n,C_n]\neq0$, the channel has only one fixed point, which is the maximum mixed state $\mathbb{I}_n/3=(P^n_1+P^n_0+P^n_{-1})/3$, and two metastable points. The convex combination of the fixed point and the two metastable points spans a two-dimensional MM holding three EMSs, which are exactly $P_\beta^m$. We note that the eigenvalues $\lambda_2$ and $\lambda_3$ of two metastable points with eigenvalues are usually different  ($|\lambda_2|>|\lambda_3|$), which makes it difficult to directly observe the three-peak feature in experiments.
As $m$ is so large that the contribution of one metastable point can be neglected ($|\lambda_3|^m\to 0$), the remaining metastable point, together with the fixed point, span a one-dimensional MM with two EMSs, which can be calculated by Eq. \eqref{1DEMS}. For Fig. \ref{data}, numerical calculations show that the two observed EMSs are $\rho_{\rm D}=P_1^n$ and $\rho_{\rm B}=(P_0^n+P_{-1}^n)/2$. 


\textbf{Weak measurements of the probe spin} 

In this section, we model the optical readout process of NV centers at room temperature. 
If $n$ photons are collected by the counter, the Kraus operator $K_n$ acting on the NV electron spin is \cite{Meinel2022}
\begin{equation}
    K_n=\sum_{\alpha=0,1}\sqrt{p(n|\alpha)}P_\alpha^e,
\end{equation}
where $P_\alpha^e$ is projector on subspace $\alpha$ and collected photon numbers obey Poisson distribution  $p(n|\alpha)=\frac{1}{n!}e^{-n_\alpha}n_\alpha^n$ with $n_\alpha$ being the average photon number obtained for state $|\alpha\rangle_n$. If $n_\alpha$ is small (about $0.05$ in our experiments), truncating to $K_0$ and $K_1$ is a good approximation with $p(1|\alpha)=n_\alpha$ and $p(0|\alpha)=1-p(1|\alpha)$.

Then the Kraus operator on the $^{14}$N bath spin when obtaining $n$ photons can be derived as 
\begin{equation}
\begin{aligned}
          W_{n}(\rho)&=\Tr_e[K_n R_2 U R_1(|0\rangle_e\langle 0|\otimes \rho) R_1^\dagger U^\dagger R_2^\dagger K_n^\dagger ] \\
            &=\sum_{\alpha=0,1}{p(n|\alpha)} M_\alpha\rho M_\alpha^\dagger,
\end{aligned}
\end{equation}
where $M_\alpha=[U_0-(-1)^\alpha e^{i\Delta \phi}U_1]$ is the Kraus operator introduced in Eq. \eqref{channel}, which corresponds to perfect projective measurements. Note that 
\begin{equation}
\begin{aligned}
             \hat\Phi'=\sum_{n}\hat{W}_n&=\sum_{\alpha=0,1}\sum_{n}p(n|\alpha)\hat{M}_\alpha\\
             &=\hat\Phi,
\end{aligned}
\end{equation}
then the channel with the set of Kraus operators $\{W_n\}_{n=0}^{\infty}$ is the same as that with $\{M_{\alpha}\}_{\alpha=0,1}$, so the analysis of the channel decomposition in the section above also applies here.  

\textbf{Measurement statistics of sequential RIMs}. 

The fixed points (or metastable points) can be observed by the measurement statistics of sequential RIMs. A single RIM includes the measurement on the probe spin with an outcome $\a\in\{0,1\}$, while for sequential RIMs we obtain a sequence of binary numbers $\{\a_1,\cdots,\a_m\}$. This sequence of measurement outcomes defines a quantum trajectory. We then look for some statistical observable which can characterize different quantum trajectories and their corresponding fixed points.

In practice, we often focus on the average of the $m$ measurement results and its expectation, i.e., $\bar{\a}=\frac{1}{m}\sum_{n=1}^m \a_n$, which in our case coincides with $f_1$  and can be related to the measurement polarization by $X=1/2-f_1$. The expectation of measurement average can be written as
\begin{equation}
    \expect{f_1} = \sum_{\a_1} \cdots \sum_{\a_m} f_1 p(\a_1,\cdots,\a_m|\rho),
\end{equation}
where the probability to get a specific sequence of measurement results $\{\a_1,\cdots,\a_m\}$ is given by
\begin{equation}\label{pniid}
    p(\a_1,\cdots,\a_m|\rho)=\brakett{\mathbb{I}|\supoper{M}_{\a_m} \cdots \supoper{M}_{\a_1}}{\rho}.
\end{equation}
It can be shown that in the asymptotic limit, i.e., when number of repetitions is large enough $m \rightarrow \infty$, $\expect{f_1}$ is only determined by fixed points, 
\begin{equation}
    \lim_{m \rightarrow \infty} \expect{f_1} = \sum_{j=1}^J c_j \expect{f_{1j}}_{*},
\end{equation}
where we assume there are $J$ fixed points, with 
\begin{equation}\label{f1fixed}
    \expect{f_{1j}}_{*} =\brakett{\mathbb{I}|\supoper{M}_1}{\rho^{j}_{\mathrm{fix}}}=p(1|\rho^{j}_{\mathrm{fix}}),
\end{equation}
where $c_j=\Tr(\rho^{j}_{\mathrm{fix}} \rho)$ amounts to the probability of obtaining $j$th fixed point given initial state $\rho$, and note the second equation indicates that the expectation of measurement average coincides with the probability of obtaining result $\a=1$ for the fixed point $\rho^{j}_{\mathrm{fix}}$ in a single RIM,
\begin{equation}\label{mprob}
    p({\a}|\rho)=\Tr(M_{\a} \rho M_{\a}^{\dagger})=\brakett{\mathbb{I}|\supoper{M}_\a} {\rho}.
\end{equation}
In other words, the distribution of the measured frequency can have $J$ peaks (without degeneracy), each centered around the expectation determined by each fixed point $\{\expect{f_{1j}}_{*}\}_{j=1}^J$ respectively. Each peak includes trajectories defined by $\{\a_1, \cdots ,\a_m\}$ with a frequency $f_1$ close to $\expect{f_{1j}}_*$, which can steer any initial state to one of the fixed point subspaces \cite{Ma2023}.

Taking into account the weak measurements, the average photon number corresponding to the $j$th fixed (metastable) point is
\begin{equation}
\begin{aligned}
         \langle n_j\rangle_*&=\sum_n\braa{\mathbb{I}}n\hat {\mathcal{W}}_n|{\rho_{\rm fix}^j}\rangle\rangle\\
         &=\sum_{n,\alpha}n p(n|\alpha) \langle f_{\alpha j}\rangle_*
\end{aligned}
\end{equation}
in which $\hat{\mathcal{W}}_n=W_n\otimes W_n^*=\sum_{\alpha=0,1}{p(n|\alpha)} \mm_\alpha$ is the superoperator of $W_n$ in the HS space.

\newpage
\,
\newpage

\onecolumngrid
\setcounter{figure}{0}
\renewcommand{\figurename}{Extended data Fig.}

\begin{figure*}
    \centering
    \includegraphics[width=18cm]{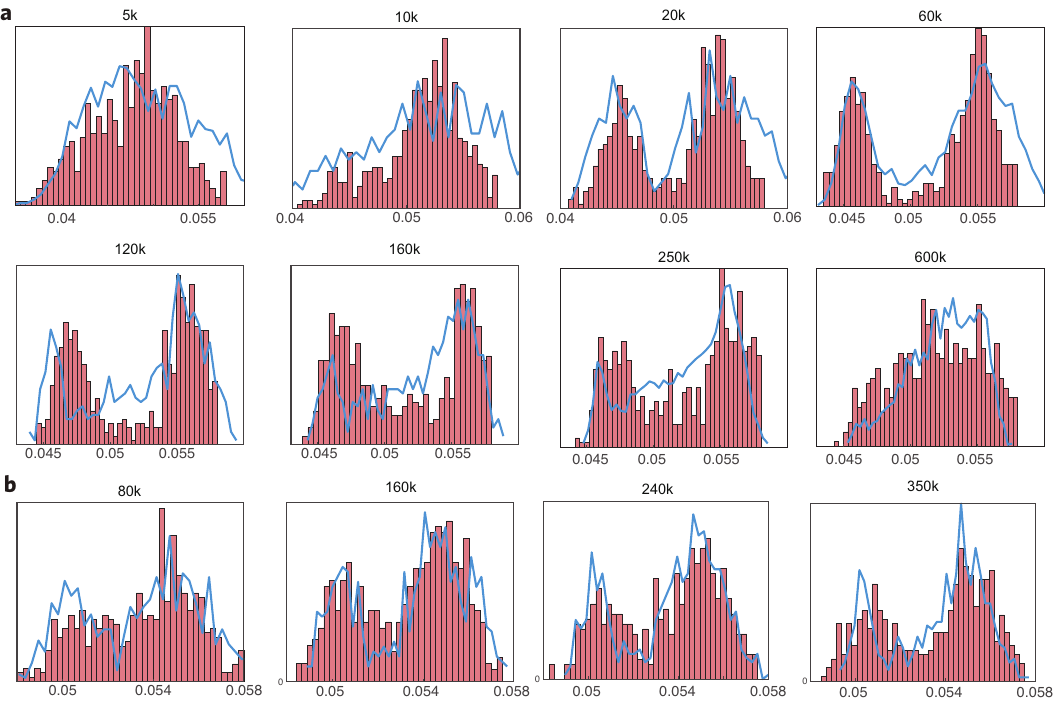}
    \caption{ \textbf{a,} Statistical results of the PL trace of the experiment with field tilt angle $7^\circ\pm 2^\circ$ and $\tau=374$ ns, data of 5 k, 60 k, 250 k, 600 k are shown in Fig. 2\textbf{c}). \textbf{b,} another experiment with the field tilt angle of $8^\circ\pm 2^\circ$ and $\tau=370$ ns, the distribution of consecutive measurements in experiment (red histogram) and simulation (blue fitting line). The simulation in \textbf{b} is performed under $\theta=7.94^\circ$, which is close to the angle in the experiment.}
    \label{fig:Ex_data1}
\end{figure*}
\newpage

\begin{figure*}
    \centering
    \includegraphics[width=18cm]{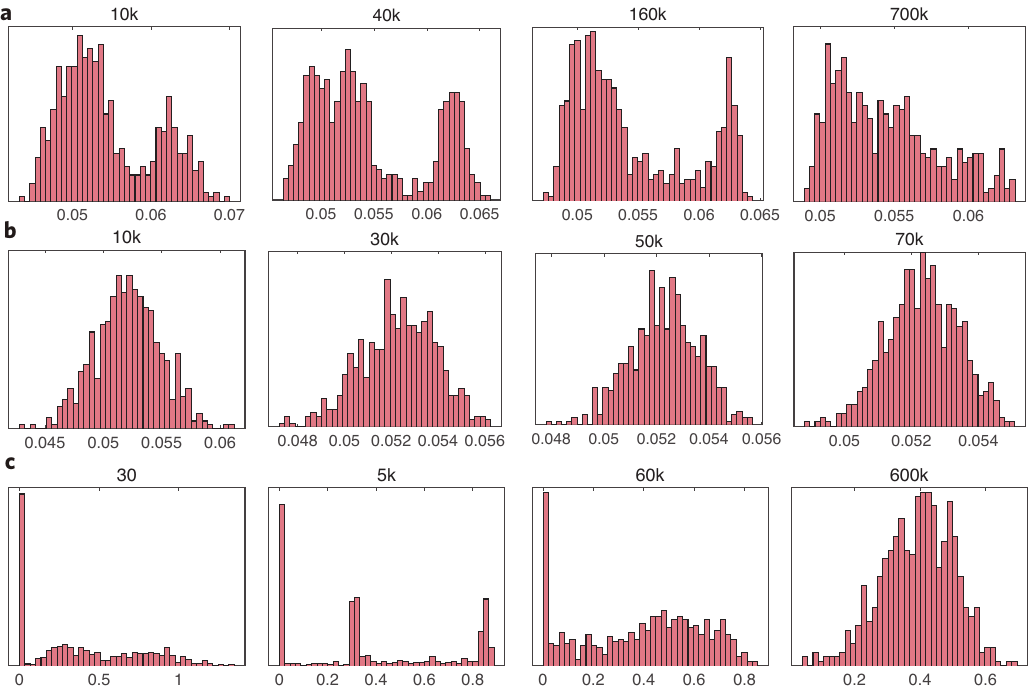}
    \caption{Histogram of the results of the Monte Carlo simulation. The parameters of \textbf{a-b} are \textbf{a} $\theta=3.5^\circ$ and \textbf{b} $\theta=15^\circ$ with weak measurement ($n_0=0.065$ and $n_1=0.049$). \textbf{c,} numerical simulation results with strong measurement ($n_0=0$ and $n_1=1$) with $\theta=8.8^\circ$}
    \label{fig:edf2}
\end{figure*}

\begin{figure*}
    \centering
    \includegraphics[width=18cm]{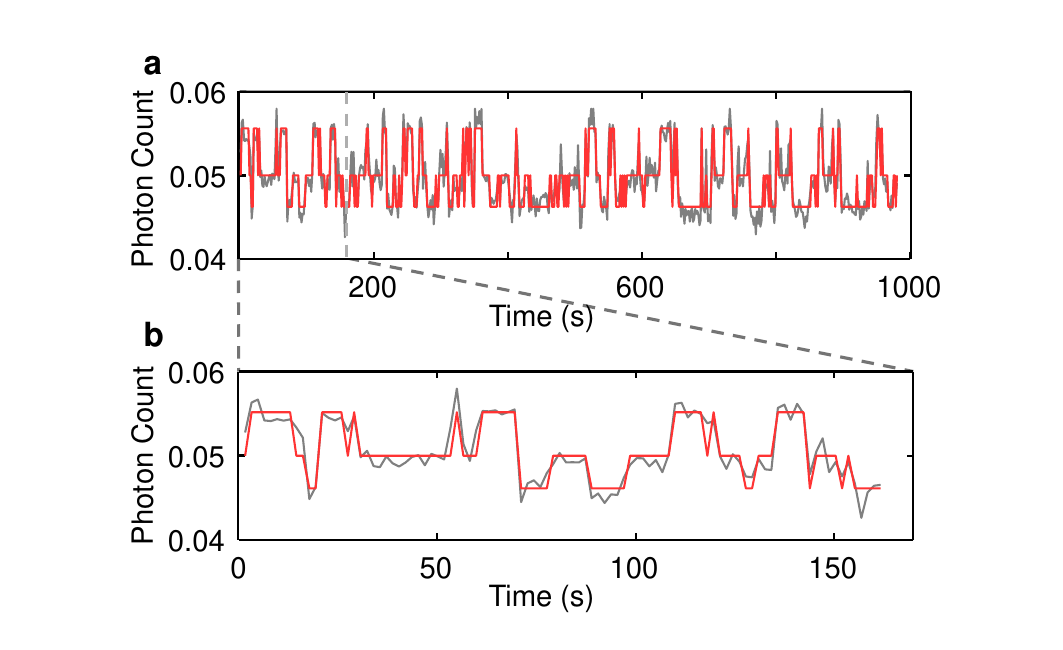}
    \caption{Triple-peak quantum jumps. 
    \textbf{a,} With a slight misalignment of the magnetic field with respect to the NV axis, the signal of triple-peak quantum jumps is  observed at the RIM duration $\tau = 350$ ns, $m$ = 60 k. The red fitting line is extracted from the HMM model.
    \textbf{b,} The first 100 points of of triple-peak quantum jumps signal.
    }
    \label{edf3}
\end{figure*}
\newpage




\clearpage

\end{document}